\documentclass[11pt, journal, onecolumn, draftclsnofoot]{IEEEtran}
\ifCLASSINFOpdf
\else
\fi

\usepackage{amsmath,amssymb,amsfonts}
\usepackage{algorithmic, adjustbox}
\usepackage{graphicx}
\usepackage{textcomp}
\usepackage{lineno}

\usepackage{amsmath,amssymb,amsfonts}
\usepackage{algorithmic}
\usepackage{graphicx}
\usepackage{textcomp}
\usepackage{amsmath,graphicx}
\usepackage{caption}
\usepackage{subcaption}
\usepackage{placeins}
\usepackage{hyperref}
\usepackage[boxruled]{algorithm2e}
\usepackage{arydshln}
\usepackage{multirow}
\usepackage{xcolor}
\usepackage{placeins}

\begin{document}
%
\title{Teaching Digital Signal Processing by Partial Flipping, Active Learning and Visualization}
%
%
%

\author{Keshab K. Parhi,~\IEEEmembership{Fellow,~IEEE}
\thanks{K. K. Parhi is with the Department
of Electrical and Computer Engineering, University of Minnesota, Minneapolis, 55455, USA e-mail: parhi@umn.edu. This paper has supplementary downloadable material available at \url{http://people.ece.umn.edu/\~parhi/.DATA/Parhi_IEEE_SPM_2021_Supplementary_Material.zip}. The materials include data and MATLAB codes for the Problems described in this paper. Please contact parhi@umn.edu for further questions about this work.}
}

%
%

\markboth{IEEE Signal Processing Magazine,~Vol.~xx, No.~xx, March~2021}%
{Parhi \MakeLowercase{\textit{et al.}}: Title of the Paper}
%



\maketitle

\begin{abstract}
Effectiveness of teaching digital signal processing can be enhanced by reducing lecture time devoted to theory, and increasing emphasis on applications, programming aspects, visualization and intuitive understanding. An integrated approach to teaching requires instructors to simultaneously teach theory and its applications in storage and processing of audio, speech and biomedical signals. Student engagement can be enhanced by engaging students to work in groups during the class where students can solve short problems and short programming assignments or take quizzes. These approaches will increase student interest in learning the subject and student engagement.
\end{abstract}

%
\IEEEpeerreviewmaketitle

\section{Introduction}
%
%
%
%
\IEEEPARstart{D}{igital} signal processing (DSP) is used in numerous applications such as communications, biomedical signal analysis, healthcare, network theory, finance, surveillance, robotics, and feature extraction for data analysis. Learning DSP is more important than ever before because it provides the foundation for machine learning and artificial intelligence.  

The DSP community has benefited tremendously from Oppenheim's views of education~\cite{oppenheim1992personal, oppenheim2006one} and from his many field-shaping text books. Teaching an engineering class in general and the DSP class in particular today are very different from that of thirty years ago. Schulman captures how classes with significant mathematical content were taught in the past~\cite{shulman2005signature}. He describes specifically how a professor teaches fluid dynamics, “He is furiously writing equations on the board, looking back over his shoulder in the direction of the students as he asks, of no one in particular, ``Are you with me?” A couple of affirmative grunts are sufficient to encourage him to continue ... This is a form of teaching that engineering shares with many of the other mathematically intensive disciplines and professions; it is not the 'signature' of engineering.” The author is right that although some instructors teach engineering this way, that is not and should not be our teaching signature. When I taught the DSP Class at the University of Minnesota using the Oppenheim-Schafer text book \cite{Oppenheim2009Discrete} in Fall-1989, my teaching signature was close to what Schulman describes. But over last three decades, my teaching signature has changed significantly. In this paper, I will describe my teaching signature as I practice today.

The objectives of teaching are three-fold: (a) teach the necessary mathematical theory and derivations, (b) introduce sufficient applications and visualize the results by programming the application, and (c) present intuitive insight about the observations from the programming experiments. Thus, signal processing is as much about listening to sounds and visualizing temporal and spectral representations as about theoretical problem solving.

My teaching signature can be described as {\em blended teaching}. I teach mostly by writing down the content in class. This helps the students write down what I am writing down; this then helps them develop the same thought process as I do when I derive these results. I then switch to power point slides and show graphs, plots, and play audio sounds to see how a signal sounds after a certain filtering operation or how different types of filters change the signal to different forms. We filter some music, some sinusoids and discuss some MATLAB code. This blended teaching keeps the students engaged. I try to assign homework problems that relate to real applications. I remind myself of the threshold concepts~\cite{meyer2003threshold}. While thresholds differ for different students, I try to cover as many if not all of these concepts. In signal processing, we have many threshold concepts. These range from myriads of math tricks to challenges in applying the same concepts to different problems. The most challenging part of teaching is to really pretend that we are not experts but novices. Then we can teach other novices more effectively.

This paper is organized as follows. Section II describes the challenges in teaching DSP. Section III describes an integrated approach to teaching where theory, applications and visualization can be taught in an integrated manner. Section IV presents blended teaching, i.e., teaching in an active learning environment using partial flipping. Section V presents metrics used in measuring effectiveness of the teaching method described in this paper.
\section{Challenges in Teaching DSP}
When I took the DSP class at the University of Pennsylvania in 1983 using the Oppenheim-Schafer text book \cite{Oppenheim1975digital}, it was taught as an advanced graduate course at that time. The DSP class is taught today as a Senior elective at most universities.

There is a desire to teach the class as a practical class where signals, sounds and images can be manipulated using DSP. This manipulation should be integrated into lecture as well as homework. One of the challenges is that the text books are rich in theory, but do not provide sufficient number of practical applications. The text book by Mitra, however, provides numerous applications related to multirate and sample-rate alteration \cite{mitra2006digital}. Because the class is often taught with emphasis on the theory, many students lose interest in taking the class. Such a class is an elective class. So to increase enrollment, students should find the class interesting and practical. We also need to train students to acquire the practical skills that will help them in their jobs in industries. However, we also need the mathematical rigor for students interested in advanced education and research. In the absence of an ideal text book, this places burden on instructors to design application examples to be covered during lectures and applications to be assigned as part of the homework. We have already seen some local success in this direction~\cite{kanna2018bringing}.

There is also a need to increase student engagement and interest. This requires instructors to deviate from traditional teaching, and adopt some form of flipped teaching where students familiarize with some content before coming to the class either by reading content or listening to video lectures~\cite{baran2016mooc}, \cite{DSP2017}. This frees up time in class where students can work together to solve theoretical or practical problems.

To increase attendance, short quizzes can be assigned during lecture. Assigning group quizzes can enhance student engagement where students in the same group can discuss and learn from each other. Thus, taking a quiz is as much about learning as about grade in the class.

Often the homework can be frustrating if the students do not learn the tricks. Students find the lectures easy, but find it harder to solve problems. Some of the tricks to solving the problems need to be taught during lecture. This requires working out some of the problems that would have been assigned as homework. Another approach is to provide solutions to problems that are similar to the homework problems. Studying these solutions will be very helpful to the students and prepare them to solve their homework problems. Same is also true for programming problems. Starter Codes for programming assignments should be provided to the students. This will help the students to solve their programming assignments. Some students have strong theoretical skills but are less inclined to solve programming problems.

Finally there is often a gap between the homework assignments and exams. Homework problems are often time-consuming and require more calculations, whereas examinations cover short problems that take less time but are thought-provoking and non-trivial. Students need to develop skills in solving problems that are similar to those in tests. The quizzes during lectures mentioned above can be very helpful to students to prepare for exams.

\section{An Integrated Approach to Teaching}
There is debate in the community about the interrelationship between innovation and education \cite{solo2019}.  This section describes examples of how signal processing can be taught more effectively via an integrated approach that emphasizes learning of the theory, application and intuition. 

\subsection{Mathematical Derivations}
Many decades ago, the entire class was spent on deriving the mathematical theory and the homework problems were mostly mathematical. Many digital signal processing homework problems involve "tricks" that were not taught in the class but students are expected to figure these out. As a student, I enjoyed figuring out these tricks; however, many students lose interest in learning DSP as they cannot figure out these tricks. Thus, there is a need to spend lecture time solving problems where the tricks are explained. This reduces the time required for all the derivations. Fortunately, students can read the text book for this part. In general the amount of lecture time used to derive theory needs to be reduced.

The threshold concepts come into play when explaining the tricks. Students may have forgotten some of the concepts. Typically, in a junior-level class on Signals and Systems, I spend a week to teach functions, scaling and shifting of functions, complex numbers and trigonometric identities. 

Another aspect of deriving theory is to explain the results first intuitively and then derive the theory. At other times, it may be easier to compute the result by MATLAB and then explain the result theoretically. This achieves two objectives: students develop practical skills and then relate the experimental result to theory. This makes the theory more relevant. As one example, I illustrate the FFT properties using Table~\ref{tab:example}. Students then verify the properties using MATLAB (see Problems 5 and 6 in Section IV). Variations of Table~\ref{tab:example} can also be used for homework or group activity.

\begin{table*}[]
\caption{Example given in class: Let the DFT of a 6-point complex sequence $(a, b, c, d, e, f)$ be another complex sequence $(A, B, C, D, E, F)$. Then complete the table below. The sequences in bold were given and students were asked to find the corresponding pairs; the sequences in red correspond to solutions.}

\resizebox{\columnwidth}{!}{\begin{tabular}{|c|c|}
\hline
${x[n]}$                                                            & ${X[k]}$                                         \\ \hline
$\mathbf{(A, B, C, D, E, F)}$                                              & \textcolor{red}{$(6a, 6f, 6e, 6d, 6c, 6b)$}                              \\ \hline
\textcolor{red}{$(A, F, E, D, C, B)$}                                                       & $\mathbf{(6a, 6b, 6c, 6d, 6e, 6f)}$                     \\ \hline
$\mathbf{(A^*, B^*, C^*, D^*, E^*, F^*)}$                                        & \textcolor{red}{$(6a^*, 6b^*, 6c^*, 6d^*, 6e^*, 6f^*)$}                        \\ \hline
$\mathbf{(A, F, E, D, C, B)}$                                              & \textcolor{red}{$(6a, 6b, 6c, 6d, 6e, 6f)$}                              \\ \hline
$\mathbf{(a^*, b^*, c^*, d^*, e^*, f^*)}$                                        & \textcolor{red}{$(A^*, F^*, E^*, D^*, C^*, B^*)$}                              \\ \hline
$\mathbf{(a, f, e, d, c, b)}$                                              & \textcolor{red}{$(A, F, E, D, C, B)$}                                    \\ \hline
\textcolor{red}{$(a^*, f^*, e^*, d^*, c^*, b^*)$}                                                 & $\mathbf{(A^*, B^*, C^*, D^*, E^*, F^*)}$                     \\ \hline
$\mathbf{(a, -b, c, -d, e, -f)}$                                           & \textcolor{red}{$(D, E, F, A, B, C)$}                                    \\ \hline
$\mathbf{(a, -f, e, -d, c, -b)}$                                           & \textcolor{red}{$(D, C, B, A, F, E)$}                                    \\ \hline
\textcolor{red}{$(a, 0, b, 0, c, 0, d, 0, e, 0, f, 0)$}                                     & $\mathbf{(A, B, C, D, E, F, A, B, C, D, E, F)}$         \\ \hline
$\mathbf{(A, B, C, D, E, F, A, B, C, D, E, F)}$                            & \textcolor{red}{$(12a, 0, 12f, 0, 12e, 0, 12d, 0, 12c, 0, 12b, 0)$}      \\ \hline
$\mathbf{(A, 0, B, 0, C, 0, D, 0, E, 0, F)}$                               & \textcolor{red}{$(6a, 6f, 6e, 6d, 6c, 6b, 6a, 6f, 6e, 6d, 6c, 6b)$}      \\ \hline
\textcolor{red}{$(A/6, 0, F/6, 0, E/6, 0, D/6, 0,  C/6, 0, B/6, 0)$}                        & $\mathbf{(a, b, c, d, e, f, a, b, c, d, e, f)}$         \\ \hline
 \textcolor{red}{$(A/12, F/12, E/12, D/12, C/12,$}& {\multirow{2}{*}{$\mathbf{(a, 0, b, 0, c, 0, d, 0, e, 0, f)}$ }}\\
    \textcolor{red}{$B/12, A/12, F/12, E/12, D/12, C/12, B/12)$} &\\
\hline
$\mathbf{(D, 0, E, 0, F, 0, A, 0, B, 0, C, 0)}$                            & \textcolor{red}{$(6a, -6f, 6e, -6d, 6c, -6b, 6a, -6f, 6e, -6d, 6c, -6b)$} \\ \hline
\end{tabular}}
\label{tab:example}
\end{table*}

\subsection{Applications and Visualization in MATLAB}
I explain several applications of the theory during lecture. In addition, I assign programming problems for applications as part of homework. For example, when describing digital filters such as low-pass, band-pass and high-pass, I take a sound or audio file and filter these with different pass bands and then hear the filtered sound. These sounds are either embedded into the powerpoint presentation or obtained from MATLAB during the class. I also provide the codes for the filtering operations. Students can use the code for solving their homework. 

In the first DSP class, I ask students to record their speech where they describe about themselves for a few minutes. They turn in their speech as part of the homework. Then, in subsequent weeks, I ask them to filter their own speech. We have also downloaded publicly available bird sounds from web sites and students use these sounds for their homework.

In another application for speech or audio compression, we compute the FFT of the signal. Then we retain low-frequency content and compute IFFT. Then we listen to the sound. This is explored in a homework problem listed below.

The MATLAB problem in Problem 1 relates to audio compression and is assigned as part of the homework. In this problem students explore the principles of audio compression where the high frequency content is discarded. The MATLAB codes for the functions $fft\_compress$ and $fft\_expand$ are provided to the students.

\textbf{Problem 1.}
\begin{enumerate}
    \item [] a) Load the audio file, referred as $x[n]$. Let $X[k]$ be its DFT. Compute $X[k]$ using FFT command.
    
    \item [] b) Compress the FFT X[k] using the $fft\_compress$ function and a percentage of compression = 10\% (0.10). This retains only the first 10\% of the spectrum.

    \item [] c) Using the compressed sound file from part (b), apply the $fft\_extract$ function to reconstruct the original audio file. Save the reconstructed audio sound file and play it. Refer to this signal as $x_1[n]$. Comment on your observations.
    
    \item [] d) Generate an error file which is the difference between the original audio file, $x[n]$, and the reconstructed audio file, $x_1[n]$. Call this error signal $e[n]$. Save the error sound file and play it. Comment on your observations.
    
    {\em Comment.} $e[n]$ contains higher frequency content of $x[n]$.
     
    \item [] e) Observe that the error signal contains frequency components in the mid band and no frequency components at low frequency. Shift left the frequency components of the error signal by $k_0$ samples and compute the IFFT of the shifted frequency-domain signal. Save the generated sound file and play it. Call this signal $x_2[n]$. Comment on your observations.
    
    {\em Comment.}
    \begin{align*}
        X_2[k] & = E[k + k_0]\\
        x_2[n] & = e[n]e^{-j\frac{2\pi}{N}nk_0}
    \end{align*}
    The signal $x_2[n]$ is complex and differs from $e[n]$ and is a modulated version of $e[n]$. Thus, if we listen to its magnitude, it will sound different from $e[n]$. 
    
    \item [] f) Multiply the signal $x_2[n]$ with a complex exponential $\left(e^{j \frac{2 \pi}{N} k_0 n}\right)$ where $k_0$ corresponds to the shift in frequency performed in part (e). Save the generated sound file and play it. Call this signal $x_3[n]$. Comment on your observations.
    
    {\em Comment.}
    \begin{align*}
        x_3[n] & = x_2[n]e^{j\frac{2\pi}{N}nk_0} ~=~ e[n]\\
    \end{align*}
    $x_3[n]$ is same as $e[n]$.
\end{enumerate}

{\em Solution.}
Solutions with MATLAB codes are provided in the Supplementary Information.  $\blacksquare$ 

I introduce practical applications while describing theoretical concepts. For example, while introducing the definition of {\em autocorrelation} of a real signal, I provide examples of photoplethysmogram (PPG) and respiration rate signals. Then I discuss how to compute the heart rate and respiration rate from these two signals using autocorrelation by looking at the zero crossings. We then compute the DFT of the signals and and verify if the frequency obtained by the autocorrelation is same as that from the DFT. This is illustrated in Problem 2 where the PPG signal is used to compute the heart rate. The respiration rate signal is not included in Problem 2 but the approach is similar. This helps connect the theoretical expression for autocorrelation to a practical application.

\textbf{Problem 2.}

 The Photoplethysmogram (PPG) signal captured from a sensor at a $100$Hz sampling frequency is provided in this problem. The length of the signal is $1024$ samples ($10.24$s). The data file, (ppg\_100hz\_1024samples.csv), for this problem is given in Supplementary Information. The data is a Photoplethysmogram (PPG) signal (ppg\_100hz\_1024samples.csv). PPG signal is used in this problem to compute the heart rate.

\begin{enumerate}
\item[] 
	
(a) Compute the 1024 point Fast Fourier Transform (FFT) of the signal and plot the absolute values of the single-sided FFT with a stem plot. Find the frequency in Hz of the highest magnitude in the FFT of PPG. Note that, the frequency corresponding to the highest magnitude represents the heart rate.

\item[] 
(b) In this part, we compute the heart rate using autocorreation. This is accomplished by finding the difference of the first and third zero crossings, which corresponds to the time period of the signal. This information is used to compute the heartrate in Hz from the PPG.

\end{enumerate}

\begin{figure}[h]
\begin{minipage}[b]{1\linewidth}
  \centering
  \centerline{\includegraphics[width=\linewidth]{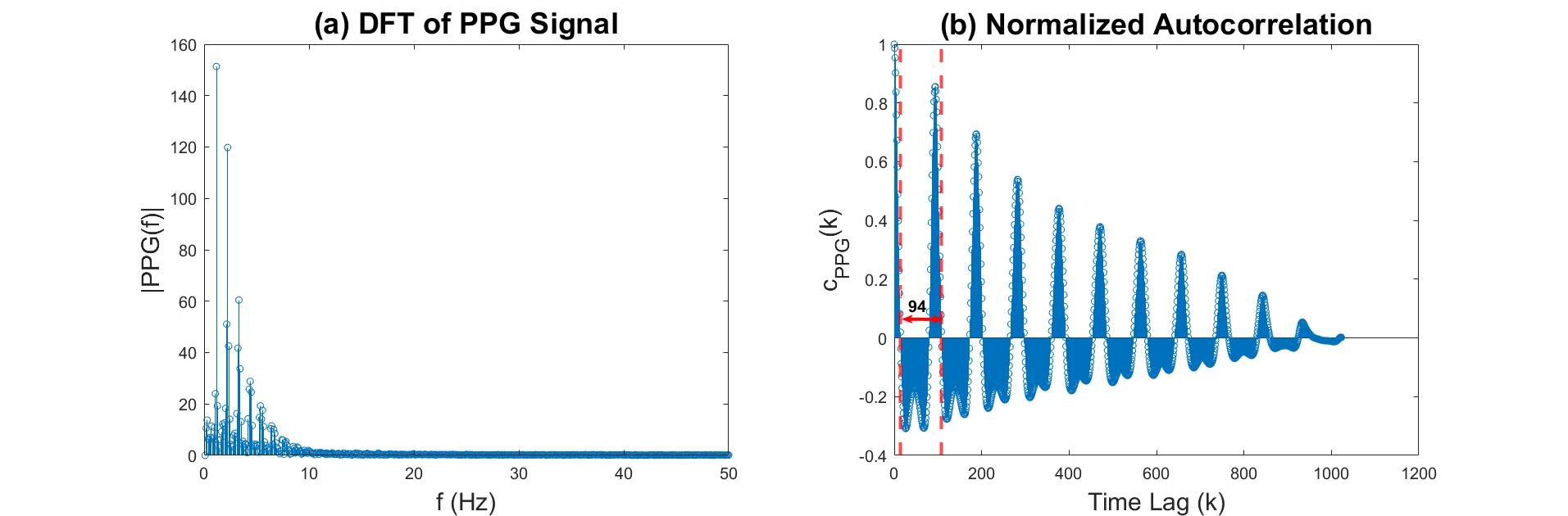}}
\end{minipage}

\caption{(a) Discrete Fourier Transform  and (b) Normalized auto correlation for the PPG signal.}
\label{fig:psd}
\end{figure}

{\em Solution.} The MATLAB code for this problem is available in the Supplementary Information. The diagram containing results from the two parts of the problem are shown in Fig.~\ref{fig:psd}. The heart rate can be estimated by 

(a) Finding the highest peak from DFT spectrum. The fundamental frequency is in $1.0742 \mathrm{Hz}$ for the PPG signal which results in heart rate of 64.45 beats per minute (bpm).

(b) Considering the interval between first and third zero crossings, a lag difference of 94 at 100Hz sampling rate = 0.94s or 63.8 bpm. Note that values from part (a) and (b) are almost same. $\blacksquare$

Fortunately large number of signals are now publicly available. These signals can be used as part of homework or class project. In my DSP class, I have used intra-cranial EEG (iEEG) signals for seizure detection from the Kaggle Seizure detection contest \cite{kaggle}. Students use the same signals for solving different programming problems assigned over many weeks and compute time-domain and frequency-domain features. These problems are described in the Appendix (see Problems 8-10). 

The role of theory and application can sometimes be interchanged. We describe an application using MATLAB first. Then we make an observation. Then we derive the theory. 

\subsection{Intuitive Insight}

It is important to explain the results from MATLAB experiments intuitively. For example, the effects of scaling and shifting a signal in the spectral domain can be quickly observed. The theory can then be explained. I assigned the problem listed below as part of a homework. 

\begin{figure}[h]
\begin{minipage}[b]{1\linewidth}
  \centering
  \centerline{\includegraphics[width=\linewidth]{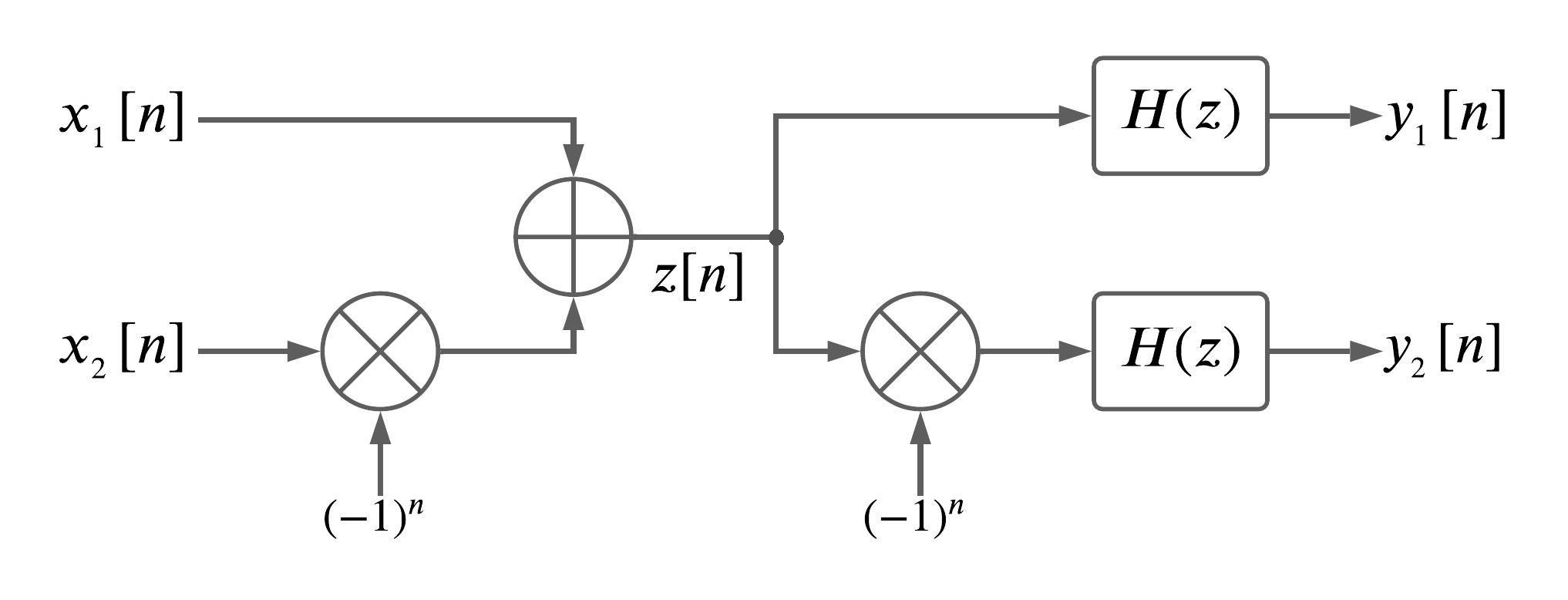}}
\end{minipage}
\caption{Signal processing System - 1 for Problem 3.}
\label{fig:problem3_1}
\end{figure}

\begin{figure}[h]
\begin{minipage}[b]{1\linewidth}
  \centering
  \centerline{\includegraphics[width=\linewidth]{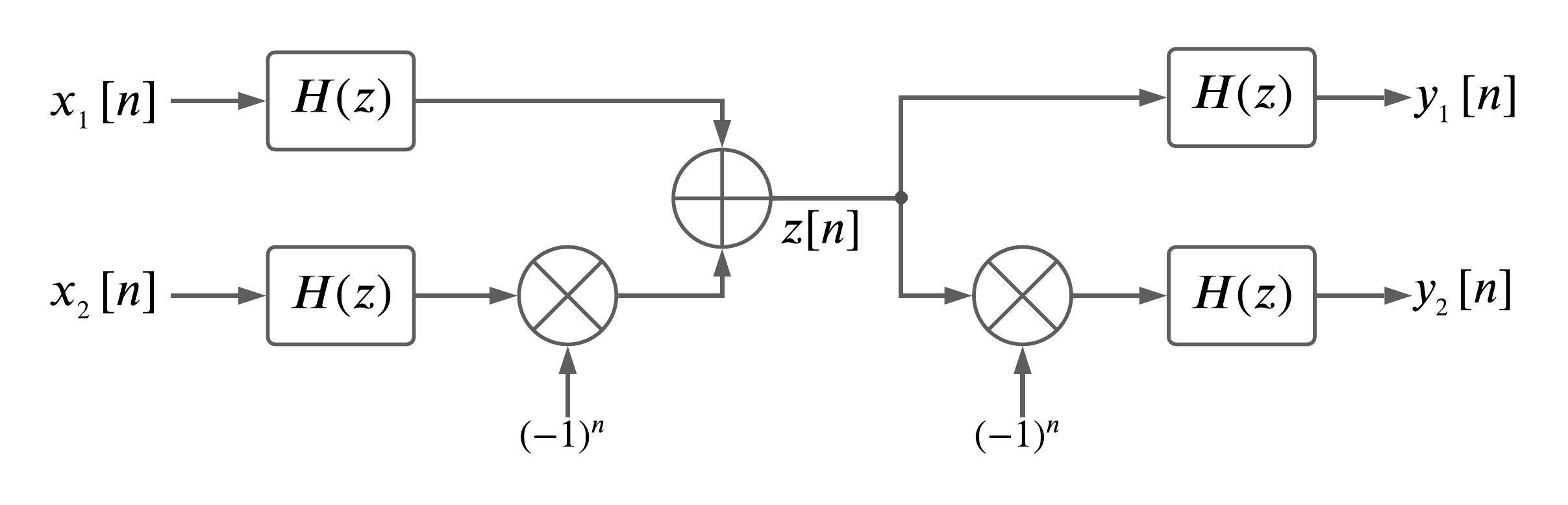}}
\end{minipage}
\caption{Signal processing System - 2 for Problem 3.}
\label{fig:problem3_2}
\end{figure}

\textbf{Problem 3.} Two signal processing systems are shown in Fig.~\ref{fig:problem3_1} and Fig.~\ref{fig:problem3_2}, where $x_1[n]$ and $x_2[n]$ are audio sounds, and $H(z)$ is a $100$th order FIR Low-Pass Filter with cut-off frequency $\frac{\pi}{2}$. For each system, load the input audio sounds and use MATLAB to obtain the sounds $z[n]$, $y_1[n]$ and $y_2[n]$ and listen to these sounds. Use {\em freqz} command to plot the spectrum of the sounds $x_1[n]$, $x_2[n]$, $z[n]$, $y_1[n]$, and $y_2[n]$ for each of the system. Compare the output signals obtained using the two DSP systems.

{\em Solution.} Multiplication of $x_2[n]$ by $(-1)^n$ results in a shift in the frequency
domain by $\pi$. Thus, the signal $z[n]$ contains the audible $x_1[n]$ along with the shifted version of $x_2[n]$ which is inaudible to the human ear. The second multiplication of $z[n]$ by $(-1)^n$ results again in a shift in the frequency
domain by $\pi$, thus making it possible to listen $x_2[n]$ at $y_2[n]$. However, $y_2[n]$ also contains the high-frequency content of $x_1[n]$. This can be avoided by band-limiting the input signals using System-2 shown in Fig.~\ref{fig:problem3_2}. $\blacksquare$

The intrigue in the above problem lies in the fact that the sound $z[n]$ does not seem to contain $x_2[n]$ whereas it is audible in $y_2[n]$. Students think that the signal $x_2[n]$ is lost and they are surprised that it can be recovered. I then explain this mathematically and intuitively. This can illustrate the basic concepts of audio steganography.

\section{Blended Teaching and Active Learning}
Almost all students today have their own laptops that they can bring to the class. Thus, it is easy for them to learn in an active learning environment where they can write short programs or solve short problems during the class. Students can learn from each other by working in groups.

\begin{figure}[h]
\begin{minipage}[b]{1\linewidth}
  \centering
  \centerline{\includegraphics[width=\linewidth]{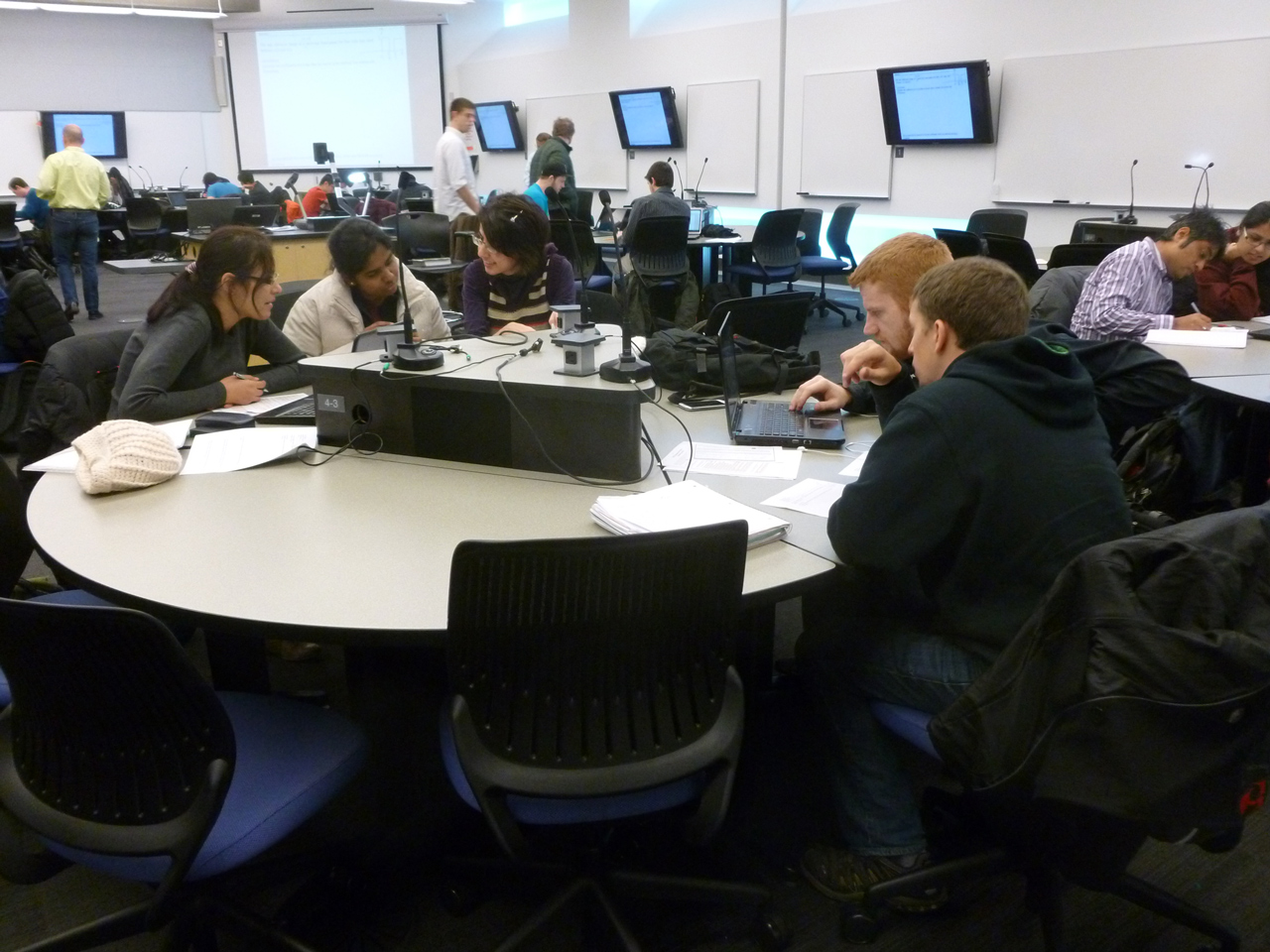}}
\end{minipage}

\caption{Active learning classroom}
\label{fig:alc}
\end{figure}

Flipped classes have been used to teach DSP effectively \cite{van2013flipping}. At the University of Minnesota, I taught the undergraduate DSP class (EE-4541) in an active learning class room in Fall-2013~\cite{parhi2013}\cite{parhi2013video}. The students sat around tables in groups of three. The classroom was equipped with a camera for the instructor and there were TV screens near each Table as shown in Fig.~\ref{fig:alc}. I used pen and paper sometimes to derive or explain theoretical results. At other times, I used powerpoint slides. Students had access to my powerpoint slides a week before the class. They were asked to review the slides before coming to class.

I assigned a quiz in the first class. The students were divided into three groups based on their performance in the quiz: top, middle and bottom. Groups consisting of three students per group were created by randomly picking one student from each of the top, middle and bottom groups. This course was taught twice per week where each class was of 75 minute duration. Out of the 75 minutes, last 15 minutes were reserved for either a group activity or a group quiz. During the group activity, the students were assigned short problems and short MATLAB assignments to work in groups. In this approach, a student who needed help could learn from another student in the group. The group quizzes also consisted of short problems and short MATLAB programming problems. The group quiz and group activity alternated from one class to another during the semester. At the end of group activity, I was able to provide intuitive insights and solutions to the problems at the end of the lecture. 

\subsection{In-Class Group Activity}
I designed the group activity problems such that students can either learn the tricks needed to solve problems or to compute the final result by MATLAB first before the theory is presented. Other problems were designed to use MATLAB to verify what was learned from theory. 

Some examples of group activity are described below.

\textbf{Problem 4}:
Consider the sinc function as follows:
\[
x[n]=\left(\frac{\sin \frac{n \pi}{4}}{n \pi}\right)
\]
Using MATLAB, plot the discrete-time Fourier transform (DTFT) of the ten signals listed below.
\begin{itemize}
    \item []a) Plot DTFT $X\left(e^{j \omega}\right)$ using Matlab.
    
    \item [] b) Let $x_{1}[n]=x[n-10] .$ Plot $X_{1}\left(e^{j \omega}\right)$ 
    
    \item [] c) Let $x_{2}[n]=x[-n] .$ Plot $X_{2}\left(e^{j \omega}\right)$
    
    \item [] d) Let $x_{3}[n]= nx[n] .$ Plot $X_{3}\left(e^{j \omega}\right)$
    
    \item [] e) Let $x_{4}[n]=e^{j \frac{n \pi}{6}} x[n] .$ Plot $X_{4}\left(e^{j \omega}\right)$ 
    
    \item [] f) Let $x_{5}[n]=(-1)^{n} x[n] .$ Plot $X_{5}\left(e^{j \omega}\right)$
    
    \item [] g) Let $x_{6}[n]=x[n]$ * $x[n] .$ Plot $X_{6}\left(e^{j \omega}\right)$
    
    \item [] h) Let $x_{7}[n]=x^{2}[n] .$ Plot $X_{7}\left(e^{j \omega}\right)$
    
    \item [] i) Let ${x_8}[n]=x[2n] .$ Plot $\mathrm{X}_{8}\left(e^{j \omega}\right)$
    
    \item [] j) Let ${x_9}[n]=\left\{\begin{array}{c}x\left[\frac{n}{2}\right], n \text { is even } \\ 0, \quad n \text { is odd }\end{array}\right.$
    
\end{itemize}

{\em Solution.}
The MATLAB codes to this problem are provided in the Supplementary Information. The students were asked to explain the discrepancy between impulse magnitudes in the MATLAB result and the theoretical result for part (d) ($\frac{201}{2 \pi}$ {\em vs.} $1$). $\blacksquare$

As another example of group activity, I asked the students to observe the properties of DFT by solving Problem 5 using MATLAB. Then they solve Problem 6 based on their observations from Problem 5. Once the students understand the properties, I derive some of these properties theoretically in class. Finally students explore use of the DFT properties shown in Table~\ref{tab:example} as part of their homework. In this Table, the sequences in red correspond to the solutions and are assigned as the homework.

\vspace{1cm}
\textbf{Problem 5}
Evaluate the following using MATLAB.

\begin{enumerate}
    \item [] a) FFT $[2,3,4,5,6]$
    \item [] b) FFT [FFT[$2,3,4,5,6$]]
    \item [] c) FFT $[2,3,4,5,6,0,0,0,0,0]$
    \item [] d) FFT $[2,3,4,5,6,2,3,4,5,6]$
    \item [] e) FFT $[2,2,3,3,4,4,5,5,6,6]$
    \item [] f) FFT $[4,5,6,2,3]$
    \item [] g) FFT $[2,-3,4,-5,6]$
    
   {\em Solution.} The FFT of a sequence can be computed using the {\em fft} command in MATLAB.  $\blacksquare$
\end{enumerate}

\vspace{1cm}
\textbf{Problem 6.}
Let $(a,b,c,d,e)\Longleftrightarrow(A,B,C,D,E)$. Write general expression for the FFT's listed below in terms of (A, B, C, D, E) based on the observations from Problem 5.

\begin{enumerate}
    \item [] a) FFT [FFT[$a,b,c,d,e$]]
    
    {\em Solution.} $[5a , 5e , 5d , 5c , 5b]$.  $\blacksquare$
    
    \item [] b) FFT $[a,b,c,d,e,0,0,0,0,0]$
    
    {\em Solution.} $[A, *, B , * , C , * , D , * , E , *]$. Here $*$ denotes interpolated values and hence cannot be expressed in terms of $A,B,C,D,E$.  $\blacksquare$
    
    \item [] c) FFT $[a,a,b,b,c,c,d,d,e,e]$
    
   {\em Solution.} $[\mathrm{A}, \mathrm{B}, \mathrm{C}, \mathrm{D}, \mathrm{E}] \left[1+e^{-j \frac{k \pi}{5}}\right]$ for $k = 0, 1, ..., 9$.  $\blacksquare$
    
    \item [] d) FFT $[a,b,c,d,e,a,b,c,d,e]$
    
    {\em Solution} $[2A , 0 , 2B, 0 , 2C , 0 , 2D , 0 , 2E , 0]$.  $\blacksquare$
    
    \item [] e) FFT $[d,e,a,b,c]$
    
    {\em Solution.} $[\mathrm{A}, \mathrm{B}, \mathrm{C}, \mathrm{D}, \mathrm{E}] e^{-j \frac{4 k \pi}{5}}$ for $k = 0, 1, ...4$. Using the property $x((n-2))_{5} \leftrightarrow e^{-j \frac{2 \pi k}{5}(2)}$.  $\blacksquare$
    
    \item [] f) FFT $[a,-b,c,-d,e]$. 
    
    {\em Solution.} Interpolation with two and a half sample delay. Here the input can be expressed as $(-1)^{n} x[n]=x[n] e^{-j \frac{2 \pi n}{5}\left(\frac{5}{2}\right)}$.  $\blacksquare$
    
    \item [] g) FFT $[a,e,d,c,b]$
    
    {\em Solution.} $\left[A^{*}, B^{*}, C^{*}, D^{*}, E^{*}\right]$. Using the property $x([-n])_{5} \leftrightarrow X^{*}[k]$.  $\blacksquare$
\end{enumerate}

\subsection{In-Class Group Quiz }
Students take group quiz once a week that lasts 15 minutes. This engages the students in the group to solve the problems together. This also reduces the pressure of taking quizzes for individual students. An example of group quiz is given below. Group quizzes help the students to be prepared for the examinations. 

\textbf{Problem 7.} Evaluate the followings. Note that these time-domain convolution problems are easier to solve in the frequency-domain.

\begin{figure}[h]
\begin{minipage}[b]{1.1\linewidth}
  \centering
  \centerline{\includegraphics[width=\linewidth]{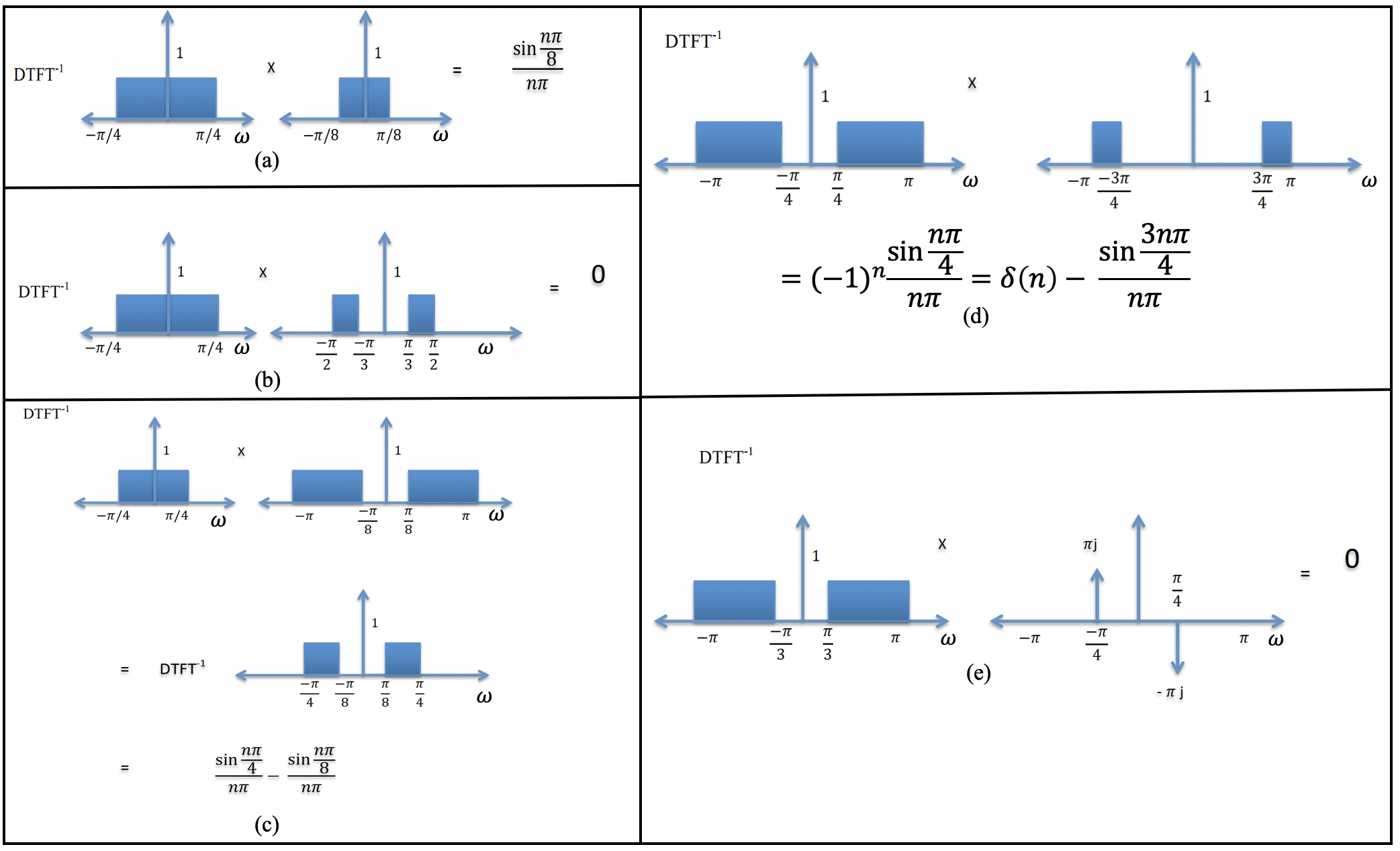}}
\end{minipage}

\caption{Problem 7 solutions.}
\label{fig:5pic}
\end{figure}

\begin{enumerate}

    \item [] a) $\frac{\sin{(n\pi/4)}}{n\pi} \ast \frac{\sin{(n\pi/8)}}{n\pi}$
    
    {\em Solution.}  Frequency-domain representation of a sinc signal is a rectangular function. The solution then involves multipying two rectangular functions and then taking an inverse Fourier transform, which is another sinc function. See Fig.~\ref{fig:5pic}(a).  $\blacksquare$

    \item [] b) $\frac{\sin{(n\pi/4)}}{n\pi}\ast(\frac{\sin{(n\pi/2)}}{n\pi}-\frac{\sin{(n\pi/3)}}{n\pi})$
    
    {\em Solution.} See Fig.~\ref{fig:5pic}(b).  $\blacksquare$

    \item [] c) $\frac{\sin{(n\pi/4)}}{n\pi}\ast(\delta[n]-\frac{\sin{(n\pi/8)}}{n\pi})$
    
    {\em Solution} Same as before; we take advantage of the fact that Fourier transform of $\delta$ function is 1. See Fig.~\ref{fig:5pic}(c).  $\blacksquare$

    \item [] d) $(\delta[n]-\frac{\sin{(n\pi/4)}}{n\pi})\ast ((-1)^n\frac{\sin{(n\pi/4)}}{n\pi})$
    
    {\em Solution.} Note, $e^{-j\pi n} = (-1)^n$. Thus the signal is shifted in the frequency-domain by $\pi$. See Fig.~\ref{fig:5pic}(d).  $\blacksquare$

    \item [] e) $((-1)^n\frac{\sin{(2n\pi/3)}}{n\pi}) \ast sin(\frac{n\pi}{4}) $ 
    
    {\em Solution.} See Fig.~\ref{fig:5pic}(e).  $\blacksquare$
\end{enumerate}

\FloatBarrier
\section{Evaluation Metrics}

\subsection{Group Comparison}

Data were collected from EE-4541 students at the University of Minnesota (UMN) in Fall-2012 (41 respondents out of 77 enrolled) and Fall-2013 (62 respondents out of 87 enrolled). The data collection was approved by the Institutional Review Board (IRB) at the UMN under the "Exempt" category. The metrics for Fall-2012 serve as the baseline for the comparison.

The two groups of students did not differ significantly on any available demographic variables, including: undergraduate-graduate status, year in the university, ethnicity, sex, age, cumulative GPA and composite ACT score. We can conclude that, as far as can be determined from the available data, the students in the two sections of EE 4541 can be validly compared to one another. 

\subsection{Outcome analyses}
The metrics used to understand the efficacy include: {\em engagement, enrichment, flexibility, effective use,
classroom/course fit, confidence}, and {\em student learning outcome} (SLO). For each measure, a number of criteria and questions were chosen for the students and they were asked to grade each question as: Strongly Agree, Agree, Disagree, and Strongly Disagree, corresponding to numerical scores of 3,2,1, and 0, respectively. A brief description of each of the metrics is presented next.

\textbf{Engage:} 
\begin{itemize}
    \item Encourages my active participation.
    \item Promotes discussion.
    \item Helps me develop connections with my classmates.
    \item Helps me develop connections with my instructor.
    \item Engages me in the learning process.
\end{itemize}

\textbf{Enrich:} 
\begin{itemize}
    \item Enriches my learning experience.
    \item Makes me want to attend class regularly.
    \item Increases my excitement to learn.
\end{itemize}

\textbf{Flexibility:}

\begin{itemize}
    \item Facilitates multiple types of learning activities.
    \item Nurtures a variety of learning styles.
\end{itemize}

\textbf{Effective Use:}

\begin{itemize}
    \item The instructor is effective in using the technology
available in the classroom for instructional purposes.
    \item The instructor is effective in using the classroom for
instructional purposes.
\end{itemize}


\textbf{Classroom/Course Fit:}

The questions asked for Classroom/Course Fit are the following  
\begin{itemize}
    \item Classroom is an appropriate space in which to hold this particular course.
    \item The in-class exercises for this course are enhanced by
the features of this classroom.
\end{itemize}

\textbf{Confidence:}

The students rate the course based on the following criteria: 
\begin{itemize}
    \item Course helps  develop confidence in working in small groups.
    \item Helps students develop confidence in analyzing.
    \item Helps student develop confidence in presenting.
    \item Helps develop confidence in writing.
    \item Improves confidence that the student can speak clearly and
effectively.
\end{itemize}

\textbf{Student Learning Outcome (SLO):}
\begin{itemize}
    \item Helps me develop professional skills that can be transferred
to the real world.
    \item Helps me to define issues or challenges and identify possible
solutions.
    \item Prepares me to implement a solution to an issue or
challenge.
    \item Helps me to examine how others gather and interpret data
and assess the soundness of their conclusions.
    \item Deepens my understanding of a specific field of study.
    \item Assists me in understanding someone else's views by imagining how an issue looks from his or her perspective.
    \item Helps me to grow comfortable working with people from
other cultures
\item Improves my confidence that I can speak clearly and
effectively.
\item Encourages me to create or generate new ideas, products,
or ways of understanding.
\item Prompts me to incorporate ideas or concepts from different
courses when completing assignments.
\item Enabled the instructor to make intentional connections
between theory and practice in this course.
\end{itemize}

\subsubsection*{Bivariate tests}
Independent-samples t-tests were conducted to compare the learning metrics of the students taught in the Fall-2013 semester using partial flipping and active learning {\em vs.} those in the Fall-2012 class with traditional learning. The results are summarized in Table~\ref{tab:et}. On all aggregated variables derived from student responses, statistically significant differences (at the $p < .05$ level or better) were found between the mean scores of the two groups favoring the Fall 2013 class (see Table~\ref{tab:et}). The group-level difference was the highest in categories of engage, flexibility and confidence. The next highest categories include classroom/course fit and student learning outcome. There is still room to improve the scores in enrich and effective use categories.

\begin{table}[ht]
\caption{Student evaluation metrics}
\label{tab:et}
\centering
\begin{tabular}{|c|c|c|c|c|}
\hline
{\color[HTML]{333333} \textbf{Variable}} & {\color[HTML]{000000} \textbf{Semester}} & {\color[HTML]{000000} \textbf{N}} & {\color[HTML]{000000} \textbf{Mean Score}} & {\color[HTML]{000000} \textit{\textbf{p}}} \\
\hline
                                         & Fall 2012                                & 41                                & 2.345                                 & 0.000                                          \\
                                         
\multirow{-2}{*}{Engage}                 & Fall 2013                                & 62                                & 2.960                                 &                                            \\
\hline
                                         & Fall 2012                                & 41                                & 2.565                                 & 0.019                                      \\
                                        
\multirow{-2}{*}{Enrich}                 & Fall 2013                                & 62                                & 2.854                                 &                                            \\\hline
                                         & Fall 2012                                & 41                                & 2.329                                      & 0.000                                          \\
                                         
\multirow{-2}{*}{Flexibility}            & Fall 2013                                & 62                                & 3.169                                      &                                            \\
\hline
                                         & Fall 2012                                & 41                                & 2.793                                      & 0.028                                      \\
                                         
\multirow{-2}{*}{Effective}              & Fall 2013                                & 62                                & 3.129                                      &                                            \\
\hline
                                         & Fall 2012                                & 41                                & 2.598                                      & 0.004                                      \\
                                         
\multirow{-2}{*}{Fit}                    & Fall 2013                                & 62                                & 3.024                                      &                                            \\
\hline
                                         & Fall 2012                                & 41                                & 2.327                                & 0.001                                      \\
                                        
\multirow{-2}{*}{Confidence}             & Fall 2013                                & 62                                & 2.672                                 &                                            \\
\hline
                                         & Fall 2012                                & 41                                & 2.497                                 & 0.007                                      \\
                                         
\multirow{-2}{*}{SLO}                    & Fall 2013                                & 62                                & 2.768                               &         \\
\hline
\end{tabular}
\end{table}


\section{Current Trends}
While MATLAB is used in many universities and industries, it is not an open-source environment. There is great interest in teaching DSP using Python or Octave as these do not require licenses. However, students have to write code from scratch for many DSP functions unlike in MATLAB where students can use numerous in-built functions. Nevertheless, using Python is more desirable as most open-source libraries for machine learning functions are written in Python. There is also growing interest in teaching DSP for embedded systems such as smart phones where students can design apps for cell phones \cite{schafer2015class}. For example, they can write DSP programs in Python for smart phones to analyze biomedical signals such as electrocardiogram. Most commercial products like smart watches already have this capability. We should create DSP Lab courses to teach application (app) design for either android or IOS operating systems to prepare students for the rapidly changing job environment.

The entire world was disrupted by COVID-19 during the initial White Paper submission of this paper (Feb. 2020) and submission of Full Paper (submitted June 2020 and revised Dec. 2020). Almost all classes in the second half of the Spring semester and the Fall semester of 2020 were taught using remote learning. This provided a challenge and opportunity to redesign various courses. Many laboratories were redesigned so that the students can perform the experiments at home. I provided lecture notes and recorded videos of the EE-4541 class to my students in Fall-2020 semester. All programming problems discussed in this paper were assigned as homework in the Fall-2020 class. Active learning is still possible using breakout rooms in a remote learning environment such as Zoom; however, it is better suited for an in-person class.

\section{Conclusion}
We argue that DSP can be effectively taught by using visualization, active learning and partial flipping. Application examples enable visualization where instructors can play different sounds and illustrate plots of time-domain and spectral-domain features. This will increase student engagement, student interest in the class, and their understanding of the subject. Use of speech, audio and biomedical signals in the class and as part of homework can connect the theory to applications and prepare students better for jobs in industries. Future DSP textbooks should include application examples and connect the theory to applications. However, instructors can use the application examples presented in this paper to supplement the text book. 

\section{Acknowledgment}
The partial flipping experiment was carried out when the author participated in the 2012-2013 Office of Information Technology (OIT) Faculty Fellowship Program at the University of Minnesota (UMN). The EE-4541 class was taught in Fall-2012 semester without flipping (baseline) and in Fall-2013 with partial flipping in an Active Learning class room. The author is most grateful to Kim Wilcox and Lauren Marsh from the UMN OIT who coordinated the program. The data collection for the OIT Faculty Fellows Program was approved by the UMN Institutional Review Board (IRB) (PI: D. Christopher Brooks, Kim Wilcox). The statistical analysis of the metrics was carried out by J.D. Walker. The evaluation reported in Section V is his contribution to this paper. All faculty fellows met multiple times during the two year program. The faculty group discussed many aspects of teaching ranging from use of apps to applying concepts of game theory~\cite{epper2012} and crowd sourcing in teaching. The author is grateful to Bhaskar Sen for his help in preparation of this paper. Zisheng Zhang and Sai Sanjay Balaji were teaching assistants for the class and developed the MATLAB codes. The author thanks Prof. M. Sabarimalai Manikandan of Indian Institute of Technology, Bhubaneshwar, India for providing the PPG application example and the PPG signal. The author's Videos from the Fall 2017 class are available at~\cite{DSP2017}.

\section*{Appendix}
This section describes programming assignments that use intra-cranial electroencephaologram (iEEG) data from the Kaggle Seizure Detection Contest organized by UPenn and Mayo Clinic \cite{kaggle}. The students were assigned one specific subject and a specific electrode/channel signal from that subject. The students were asked to extract various time-domain and frequency-domain features and comment on suitability of these features for discriminating ictal (during seizure) and interictal (baseline) clips to detect seizures. The sample solutions provided in the Supplementary Information for these problems use the training data from channel 1 for the EEG clips from Dog 1, which has a total of 596 clips of which 178 are ictal, i.e., these correspond to seizures. Each clip is a one second recording.

\textbf{Problem 8.} This problem explores time-domain signal processing. Extract and plot the eight time-domain features listed below for the assigned subjects. Use {\em stem ()} command to plot the value of each feature for all the clips. Observe the plots and comment whether the given feature could be used to detect seizures. 
\begin{enumerate}
    \item Measures of central Tendency - Arithmetic mean, median and mode.
    \item Energy: The energy for a sequence $x[n]$ of length $N$ is given by
    $$ E = \sum_{n=1}^{N} |x[n]|^2 $$
    \item Total length of the curve (sum of distances between successive points).
    $$ L = \sum_{n=2}^{N} |x[n] - x[n-1]| $$
    \item Hjorth parameters
    $$ Activity(x(t)) = var\{x(t)\} $$
    $$ Mobility(x(t)) = \sqrt{\frac{var\{\frac{d}{dt}x(t)\}}{var\{x(t)\}}}$$
    $$ Complexity(x(t)) = \frac{Mobility(\frac{d}{dt}x(t))}{Mobility(x(t))}$$
\end{enumerate}

{\em Solution.}
The MATLAB codes for this problem are provided in the Supplementary Information. The results varied based on the assigned subject and electrode. In many cases, the energy, total length and Hjorth activity were good indicators of seizure. $\blacksquare$

\textbf{Problem 9.} In this problem, we will generate an analytic signal from the EEG signal. The analytic signal is a complex signal whose real part is the signal itself and the imaginary part is the signal filtered by the Hilbert transform filter. Using this, we generate two new features listed below. Plot the features using {\em stem ()} command and comment on whether the given feature could be used to detect seizures. 

\begin{enumerate}
   \item \underline{Mean Instantaneous Amplitude of Delta band}: Filter the original EEG clips using a band-pass filter to obtain the signal in the delta band [1 Hz - 4 Hz]. Using Hilbert Transform, obtain the discrete-time analytic signal (complex-valued), the magnitude of which provides the instantaneous amplitude. Use its average as the feature.
    \item \underline{Mean Instantaneous Frequency of Alpha band}: Filter the original EEG clips using a band-pass filter to obtain the alpha band [8 Hz - 12 Hz] signal. Using Hilbert Transform, obtain the discrete-time analytic signal (complex-valued), the angle of which provides the instantaneous phase. The derivative of the unwrapped instantaneous phase scaled by the sampling frequency yields the instantaneous frequency. Use its average as the feature.
\end{enumerate}

{\em Solution.}
The MATLAB codes for this problem are provided in the Supplementary Information. The results varied based on the assigned subject and electrode. In many cases, the Mean Instantaneous Amplitude of delta band is a good indicator of seizure. $\blacksquare$

\textbf{Problem 10.} We explore the three methods listed below to observe the Power Spectral Density (PSD) of the EEG clips and find out if PSD is a useful feature for seizure detection.
\begin{enumerate}
    \item \underline{Spectrogram}: 
     Combine all the clips (ictal followed by interictal) to form a single time series. Use spectrogram command with a window of 100 sample segments and 80 samples overlap to view the frequency spectrum. Show the output as a surface plot with time on X-axis, frequency on Y-axis and Spectrum (in dB) along the third axis.
    \item \underline{Welch power spectral density estimate}: 
    Combine the ictal clips and interictal clips separately to form two different time-series. Using {\em pwelch} command, obtain and plot the PSD estimate of the two signals on the same graph for normalized frequency range of $[0 - \pi]$. Use smoothdata() function to smoothen the plot and identify the normalized frequency range that has maximum difference between the PSD estimates of ictal and interictal series.
    \item \underline{Average power spectral density as a feature}: 
    Using {\em stem ()} command, plot the average of the PSD estimate obtained using {\em pwelch} command for each clip. Observe and comment whether this feature could be used to detect seizure.
\end{enumerate}

{\em Solution.}
The MATLAB codes for this problem are provided in the Supplementary Information. The results varied based on the assigned subject and electrode. In many cases, the PSD estimate is a good indicator of seizure. $\blacksquare$


%





\ifCLASSOPTIONcaptionsoff
  \newpage
\fi



%

\FloatBarrier
\bibliographystyle{IEEEbib.bst}
\bibliography{sample.bib}

\textbf{Keshab K. Parhi } (parhi@umn.edu) received the B.Tech. degree from the
Indian Institute
of Technology (IIT), Kharagpur, India, in 1982; the M.S.E.E. degree from the
University of Pennsylvania, Philadelphia, in 1984; and the Ph.D.
degree in EECS from the University of California, Berkeley, in 1988. He has
been with the University of Minnesota, Minneapolis, since 1988, where
he is currently Distinguished McKnight University Professor and Edgar
F. Johnson Professor of Electronic Communication in the Department of
Electrical and Computer Engineering. He has published over 650 papers,
is the inventor of 31 patents, and has authored the textbook VLSI
Digital Signal Processing Systems (Wiley, 1999). His current research
addresses VLSI architecture design of machine learning systems,
hardware security, data-driven neuroscience and molecular/DNA
computing. Dr. Parhi is the recipient of numerous awards including the
2017 Mac Van Valkenburg award and the 2012 Charles A. Desoer Technical
Achievement award from the IEEE Circuits and Systems Society, the 2003
IEEE Kiyo Tomiyasu Technical Field Award, and a Golden Jubilee medal
from the IEEE Circuits and Systems Society in 2000. He served as the
Editor-in-Chief of the IEEE Trans. Circuits and Systems, Part-I during
2004 and 2005. He is a Fellow of the IEEE (1996), a
Fellow of the American Association for the Advancement of Science (2017), a fellow of the Association for Computing Machinery (2020) and a Fellow of the National Academy of Inventors (2020).

\end{document}